\documentclass[aps,twocolumn,superscriptaddress,altaffilletter,lengthcheck,
tightenlines,showpacs,showkeys]{revtex4}

\newcommand{\ben}{\begin{eqnarray}}
\newcommand{\een}{\end{eqnarray}}
\newcommand{\n}{\label}

\newcommand{\ga}{\gamma}
\newcommand{\ro}{\rho}

\usepackage{graphicx}

\usepackage{epstopdf}

\usepackage[dvipdf]{epsfig}
\usepackage{color}
\usepackage{multirow}

\begin{document}
\title{Power spectra in extended tachyon cosmologies}

\author{Iv\'an E. S\'anchez G.}\email{isg.cos@gmail.com}
\affiliation{Departamento de Matem\'{a}tica, Facultad de Ciencias Exactas y Naturales, Universidad de Buenos Aires and IMAS-CONICET, Ciudad Universitaria, Pabell\'on I, Buenos Aires  1428, Argentina}
\author{Osvaldo P. Santill\'an}\email{firenzecita@hotmail.com}
\affiliation{CONICET-Instituto de Investigaciones Matem\'aticas Luis Santal\'o, Universidad de Buenos Aires, Ciudad Universitaria, Pabell\'on I, Buenos Aires 1428, Argentina}

\date{\today}
\bibliographystyle{plain}

\begin{abstract}
In the present work  the power spectrum of a particular class of tachyon fields is compared with the one corresponding to a cosmological constant model. This is done for different barotropic indexes $\ga_0$ and the background space time is assumed to be of the spatially flat Friedmann-Robertson-Walker type. The differential equation describing the perturbations is solved numerically and the power spectrum at the scale factor value $a=1$ is plotted for each case. The result is that the power spectrum of the standard tachyon field differs in many magnitude orders from the $\Lambda$CDM. However, the one with $\ga_0=1.91$, which corresponds to a complementary tachyon field, coincides fairly well with the concordance model. Therefore, we conclude that the perturbed solutions constitute an effective method to distinguish between the different $\ga_0$ values for the tachionization $\Lambda$CDM model and the fiducial model. The Statefinder parameters $\{r, s\}$, measuring the deviations of the analysed model from the concordance model, are also explicitly calculated. Our result suggest that, depending on the value of $\gamma_0$, these models can explain the observed expansion history or the perturbation power spectrum of the universe, but they may have problems in describing both features simultaneously.
\end{abstract}
\vskip 1cm



\date{\today}
\bibliographystyle{plain}

\maketitle

\section{Introduction}

The vast majority of cosmological tests are in agreement with the simplest type of dark energy density, the cosmological constant, indicated with $\Lambda$, so the standard cosmology is usually denominated as $\Lambda$CDM model or the concordance model. Despite its striking observational success, the present value of the cosmological constant is largely unappealing from the theoretical point of view. The $\Lambda$CDM suffers  from a variety of problems that range from quantum gravity considerations to fine tuning and coincidence problems \cite{turner}-\cite{Padma}.

The standard concordance model is well studied. The formation of dark matter structures for this model is relatively well understood, thanks to both the linear perturbation theory and numerical cosmological simulations \cite{Rudd}-\cite{Heitmann}.

Cosmological simulations are one of the most powerful tools for the research of the non-linear evolution of large-scale structures of the Universe. The study of the power spectrum gives a wealth of information about the matter density field, and they are reflected in several measurements such as cosmological weak lensing.

In the recent years some attention to tachyon driven cosmologies was paid \cite{TaC}. Some reasons are that such cosmologies may play an important role in inflationary models as well as in the present accelerated expansion, simulating the effect of the dark energy, \cite{Pad}-\cite{LuisTaq2}, depending upon the form of the tachyon potential \cite{Pad}-\cite{Abramo}, \cite{Bagla}-\cite{Gilberto}. These scenarios are inspired in some string cosmology models \cite{TaS}.

In references \cite{LuisTE} and \cite{ISG} there are considered three classes of tachyons, which are classified in terms of the value of their barotropic index  $\gamma_0$ (see below). On the same way, in \cite{LuisTaq2} it is shown that, in the limiting case of small $\Lambda$ the standard and complementary tachyon fields defined in that reference tend to the $\Lambda$CDM model.

The tachyonization  cosmological models described above constitute a large class of scenarios containing the $\Lambda$CMD as a particular limit. It is natural to search for observable effects allowing to distinguish them from the $\Lambda$CMD. One of such effects is the power spectrum of the gravitational potential of the primordial fluctuations for tachyon cosmologies. These perturbations are responsible for the large structure scale of the universe and, due to the non linear nature of the Einstein equations, an appreciable deviation between the two models may be observed.

An important cosmological diagnostic pair $\{r, s\}$ called Statefinder, was introduced by \cite{Sahni}. The Statefinder is
a geometrical diagnostic and allows us to characterize the properties of dark energy in a model independent manner. The Statefinder is dimensionless quantity constructed in terms of the scale factor of the Universe and its time derivatives only. This cosmological diagnostic will be applied below to the model under consideration.

The aim of the present work is to compare the power spectrum between the concordance and the taquion models mentioned above. The organization goes as follows. In section 2 the main aspects of the power spectrum of the primordial fluctuations in Friedmann-Robertson-Walker (FRW) cosmologies are briefly discussed, and the differential equation that describes the perturbations is derived. In section 3  the three classes of tachyons in terms of their barotropic index is reviewed, together with their defining equations. In section 4 the power spectrum of primordial fluctuations for these models is characterized by numerical methods and it is compared to the $\Lambda$CMD model. Section 5 contains the  discussion of the obtained results. In section 6, we applied the statefinder diagnostic pair (dubbed statefinder)  to these models.

\section{Cosmological scalar perturbations}
The small scalar perturbations in a spatially flat Friedmann-Robertson-Walker scenario in the conformal Newtonian gauge are represented by the line element \cite{Ma}
\begin{equation}
\label{ds}
d\emph{s}^2=\emph{a}(\eta)\left[(1+2\Phi)d\eta^2-(1-2\Psi)\delta_{ij}dx^{i}dx^{j}\right],
\end{equation}
where $\emph{a}(\eta)$ is the scale factor as a function of the conformal time $\eta$. The last is defined by $d\eta=dt/a$, up to an additive constant. In this gauge, the variables $\Phi$ and $\Psi$ are gauge invariant. In the situations for which  the spatial part of the energy-momentum tensor is diagonal, it follows that $\Phi=\Psi=\phi$, with $\phi$ the Newtonian potential. So, there remains only one free metric perturbation variable $\phi$. As can be seen from (\ref{ds}), the gauge invariant quantities $\Phi$ and $\Psi$ can be interpreted as the amplitudes of the metric perturbations in the conformal-Newtonian coordinate system.

In the following, a perfect fluid with an equation of state which, for each fluid species, takes a barotropic form $p=(\gamma_0 -1)\rho$ will be considered. The anisotropic stress leads to nondiagonal space-space components of the energy-momentum tensor and vanishes for a perfect fluid. In this situation $\phi=\psi$ in Eq. (\ref{ds}), as remarked above. The general from of the classical equations which describe the evolution of small perturbations in a hydrodynamical universe \cite{Muk}, can be written in the following form
\begin{equation}
\label{u}
u''-c_{s}^{2}\nabla^{2}u-\frac{\theta''}{\theta}u=\aleph.
\end{equation}
In the equation above the prime ' denotes the derivative with respect to the conformal time and $c_{s}^{2}$ is the speed of sound of the fluid. For the present work $\aleph=0$ since it is a quantity that depends on the entropy perturbation and we will be focused with adiabatic perturbations only. In this regime the equation (\ref{u}) becomes homogeneous. In addition, the variables $u$ and $\theta$ are the Mukhanov-Sasaki variables defined through the following relations
\begin{equation}
\label{vM}
u=\frac{\Phi}{4\pi G\sqrt{\ro_{0}+p_{0}}}, \qquad \theta=\frac{1}{a}\left(\frac{\ro_0}{\ro_0+p_0}\right)^{1/2}.
\end{equation}
By rewriting Eq. (\ref{u}) in such a way that $u$ is a function of the cosmic scale factor $a$ and by considering a plane-wave solution of the form $u_\textbf{k}\exp (-i\textbf{k.x})$, the equation (\ref{u}) can be cast in the form
\begin{equation}
\label{Eu}
a^{2}\frac{d^{2}u_\textbf{k}}{da^{2}}+a\left(2+\frac{\dot{H}}{H^{2}}\right)\frac{du_\textbf{k}}{da}+\frac{c_{s}^{2}k^{2}}{a^{2}H^{2}}u_\textbf{k}+\frac{\dot{H}}{H^2}u_\textbf{k}=0,
\end{equation}
where $H=\dot{a}(t)/a(t)$ is the Hubble parameter and the dot denotes the derivative with respect to time $t$.

\section{Tachyon Field Equations and $\Lambda$CDM}

The tachyon field of \cite{LuisTaq2}, \cite{LuisTE} an scalar field $\phi$ described by the addition of the following lagrangian
\begin{equation}
\label{taclag}
L=V(\phi)\sqrt{1-\partial_\mu \phi \partial^\mu\phi},
\end{equation}
to the Einstein one with non zero cosmological constant $\Lambda$. Here $V (\phi)$ is the self interaction potential \cite{LuisTaq2} for the tachyon. The background energy density and pressure of the tachyon condensate, for a flat FRW cosmology, are given by
\begin{equation}
\rho_\phi=\frac{V}{\sqrt{1-\dot{\phi}^2}}, \qquad \quad p_\phi=-V\sqrt{1-\dot{\phi}^2},  \label{roE}
\end{equation}
respectively. The Friedmann and the standard conservation equations become
\begin{equation}
\label{A}
3H^2=\Lambda+\frac{V}{\sqrt{1-\dot{\phi}^2}},
\end{equation}
\begin{equation}
\n{kg}
\ddot\phi+3H\dot\phi(1-\dot{\phi}^2)+\frac{1-\dot{\phi}^2}{V}\frac{dV}{d\phi}=0.
\end{equation}
respectively, where $H=\dot{a}(t)/a(t)$ is the Hubble parameter and $a(t)$ is the cosmic scale factor and the dot denotes derivative with respect to the time variable $t$. The equation of state for each fluid specie takes a barotropic form $p=(\gamma_0 -1)\rho$, where $\ga$ is the barotropic index,
\begin{equation}
\n{gs}
\ga=\dot\phi^2.
\end{equation}
with $0<\ga<1$ for Eqs. (\ref{roE}). The sound speed is $c^{2}_{s}=1-\ga>0$, and by use of (\ref{gs}), we can write
\begin{equation}
\n{sos}
c^{2}_{s}=1-\dot\phi^2.
\end{equation}

For the following tachyon fields potential
\begin{equation}
\label{Pl}
V(\phi)=\frac{\Lambda\sqrt{1-\ga_0}}{\sinh^{2}\frac{\sqrt{3\ga_0\Lambda}}{2}\phi},
\end{equation}
some exact solutions of the field equation (\ref{kg}) can be found, by assuming a linear dependence of the tachyon field with the cosmological time of the form
\begin{equation}
\label{Ft}
\phi=\phi_{0}t, \qquad \dot{\phi}^{2}=\phi_{0}^{2}=\ga_0,
\end{equation}
which is consistent with Eq. (\ref{gs}). For this anzatz the Hubble parameter and its derivative are given by
\begin{equation}
\label{HL}
H=\sqrt{\frac{\Lambda}{3}}\coth\frac{\sqrt{3\ga_{0}^{2}\Lambda}}{2}t, \qquad  \dot{H}=-\frac{\ga_0\Lambda}{2}\frac{1}{\sinh^{2}\frac{\sqrt{3\ga_{0}^{2}\Lambda}}{2}t}.
\end{equation}
which, after the integration gives the following cosmic scale factor
\begin{equation}
\label{aL}
a=a_0\left[\sinh\frac{\sqrt{3\ga_{0}^{2}\Lambda}}{2}t\right]^{2/3\ga_0}.
\end{equation}
It is customary to set $a_0=1$ for the actual scale factor. The Eqs. (\ref{HL}) for the Hubble parameter and the derivative can be written in terms of the scale factor as
\begin{equation}
\label{HLa}
\frac{\dot{H}}{H^{2}}=-\frac{3}{2}\ga_0\frac{1}{1+a^{3\ga_0}}  \qquad  a^{2}H^{2}=\frac{\Lambda}{3}a^2\frac{1+a^{3\ga_0}}{a^{3\ga_0}}.
\end{equation}

The formulas just derived are valid for the standard tachyon. Nevertheless there exists  two new kinds of extended tachyon fields \cite{LuisTE}, \cite{ISG}. In the last of those references it is shown that these two new types of tachyons could be derived from the standard tachyon field $(0<\ga<1)$ by use of a symmetry argument \cite{FIS}. In the following we will consider the standard and the complementary $(1<\ga)$ tachyon field.

The complementary tachyon field $\phi_c$ is characterized by $1<\dot{\phi}²_c=\ga_0$ and its potential can be deduced from the standard tachyon under the replacement $1-\ga_0\rightarrow-(1-\ga_0)$ in Eq. (\ref{Pl}), resulting in
\begin{equation}
\label{Plc}
V(\phi_c)=\frac{\Lambda\sqrt{\ga_0-1}}{\sinh^{2}\frac{\sqrt{3\ga_0\Lambda}}{2}\phi_c}.
\end{equation}
By assuming again a linear dependence of the tachyon field with the cosmological time, Eq. (\ref{Ft}),  we can get the same Hubble parameter and its derivative, Eqs. (\ref{HL}) with $1<\ga_0$. Note that, under these conditions and by use of Eq. (\ref{sos}), the sound speed becomes $c^{2}_{s}=\ga_0-1$.

At first sight the energy density (\ref{roE}) becomes 
$$
\rho_\phi=\frac{V}{\sqrt{1-\gamma_0}}
$$
and seems to have a divergence when $\gamma_0\to 1$. Note however, that the potential (\ref{Pl}) goes to zero in this limit, and the density $\rho_{\phi}$ remains finite. Thus the singularity $\gamma_0\to 1$ is in fact an avoidable one.

\begin{figure}[hbt]
\begin{center}
\includegraphics[width=8cm]{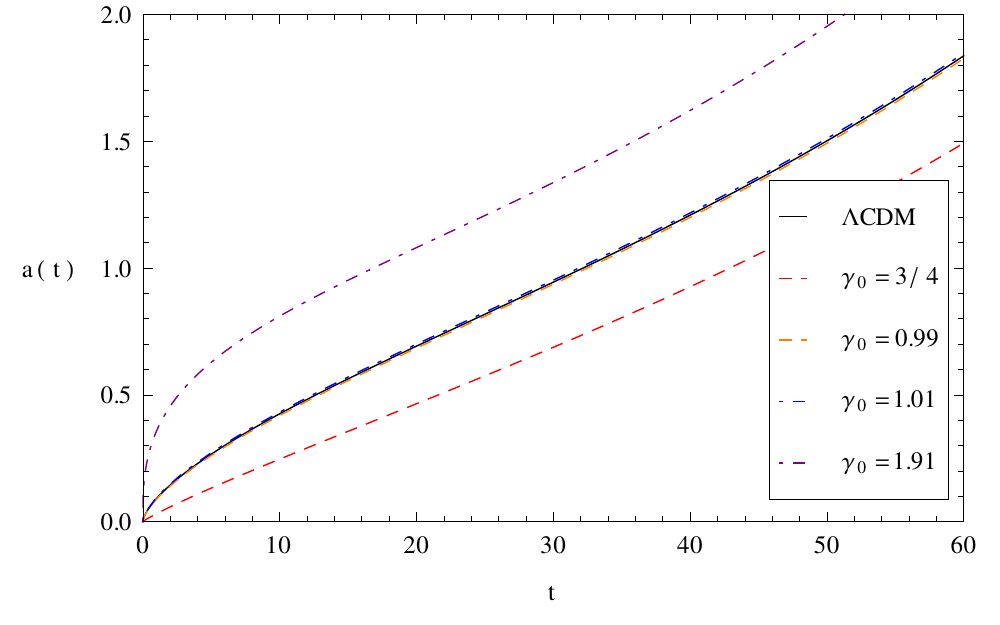}
\caption{\scriptsize{Plot of cosmic scale factors as functions of time for the standard ($\ga_0=3/4$ and $\ga_0=0.99$) and the complementary ($\ga_0=1.01$ and $\ga_0=1.9$) tachyon fields.}}
\label{F1}
\end{center}
\end{figure}

In Fig. \ref{F1} the scale factor $a$ is plotted as a function of $t$ for the $\Lambda$CDM  model and the standard and complementary tachyon field with different  $\gamma_0$ values. The standard and complementary tachyon fields for values of $\gamma_0\approx1$ ($\ga_0=0.99$ and $\ga_0=1.01$) are very close to the $\Lambda$CDM model. For these two values one infers from Fig. \ref{F1} that the cosmic scale factors increase with time, and there is no sensible difference between the solutions concerning the time evolution of the scale factors with the concordance model. The two tachyon fields here behave as matter fields which are responsible for the decelerated regime.

The standard tachyon field with $\ga_0=0.75$ plays the role of an inflationary field which decays into matter. The analyses of Fig. \ref{F1} shows that the evolution in time of the scale factor is less accentuated than that for the concordance model.

On the other hand, the complementary tachyon field with $\ga_0=1.91$ that is close to represents stiff matter ($\ga_0=2$), the increase with time of $a$ is much more accentuated than the one for the $\Lambda$CDM model. For a detailed analysis see \cite{LuisTaq2}.

\section{Power spectrum analysis}
We now proceed to present our results. Our aim is to compare the power spectrum results obtained for the tachyonization model for different values of $\gamma$ with the fiducial $\Lambda$CDM cosmology.

In the following, we shall always refer to the dimensionless power spectrum $\Delta^{2}(k,a)$ \cite{MukLibro}, defined as
\begin{equation}
\label{DPS}
\Delta^{2}_{\phi}(k,a)=16\pi^{2}(\rho+p)\mid u_{\textbf{k}}(a)\mid^{2} k^{3},
\end{equation}
with $u_{\textbf{k}}$ is the Fourier amplitude of the Mukhanov variable $u$, which is described by the Eq. (\ref{vM}). It is therefore important to set up the initial conditions for the simulation. The following initial conditions for the variable $u_\textbf{k}$ \cite{Muk}, \cite{MukLibro}
\begin{equation}
\label{CI}
u_{\textbf{k}}(a)=-\frac{i}{\sqrt{c_s}k^{3/2}}   \qquad   \frac{du_{\textbf{k}}}{da}(a)=\frac{\sqrt{c_s}}{a^{2}Hk^{1/2}},
\end{equation}
will be chosen. The reason is these are the  minimal fluctuations allowed by the uncertainty relations. According to the current inflationary theory, the primordial quantum fluctuations  are responsible for the large scale structure formation. This election is the one which minimize the  energy of these fluctuations.

We investigated the power spectrum for the tachyonization model solving Eq. (\ref{Eu}) numerically with the initial conditions (\ref{CI}). These power spectrum was compared with the fiducial $\Lambda$CDM model corresponding to the following cosmological parameters: $\Omega_{m0}=0.30$ for the total matter content, a cosmological constant contribution specified by $\Omega_{\Lambda 0}=1-\Omega_{m0}$ and the Hubble constant $H_0=71km
s^{-1} Mpc^{-1}$. These parameters are in agreement with the ones obtained in WMAP9 \cite{WMAP9} and the Planck mission \cite{Planck2013}.

\begin{figure}[hbt]
\begin{center}
\includegraphics[width=8cm]{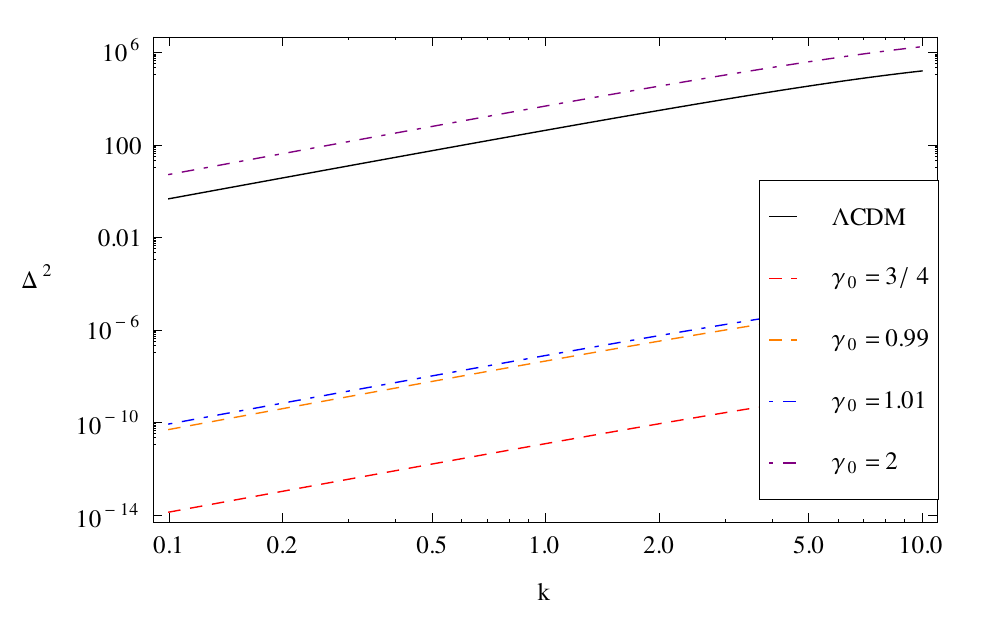}
\caption{\scriptsize{Plot of the dimensionless power spectrum as function of $k$ at $a=1$, for standard ($\ga_0=3/4$ and $\ga_0=0.99$) and complementary ($\ga_0=1.01$ and $\ga_0=2$) tachyon fields.}}
\label{F2}
\end{center}
\end{figure}

The Figure \ref{F2} describes the $a=1$ dimensionless power spectrum Eq. (\ref{DPS}) for the tachyonization of the $\Lambda$CDM model, for the standard and the complementary tachyon field with $\ga_0=3/4$, $\ga_0=0.99$ and $\ga_0=1.01$, $\ga_0=2$ respectively, and for the concordance model. These curves are the numerical solution of the Eq. (\ref{Eu}) with the corresponding initial conditions (\ref{CI}) for the above cases.

We should remark that the model with $\gamma=1$ is the one which better imitates the expansion history of the universe corresponding to the fiducial model. However, their power spectrum differs considerably. 
The most evident feature about this figure is the large discrepancy between the power spectrum for all the curves. As it can be seen from these results, there is a strong dependence of the power spectrum with the barotropic index $\ga_0$. For example, by changing from a barotropic index $\ga_0=3/4$ to $\ga_0=0.99$ the dimensionless power spectrum varies in several orders of magnitude. Comparison with the $\Lambda$CDM model reveals that all the curves are strictly below the concordance model, with the exception of the one with $\ga_0=2$. The last  one is strictly above. Figure \ref{F2} also shows that the lowest curve corresponds to the one with smallest barotropic index ($\ga_0=3/4$) and increases when $\gamma_0$ gets larger.

\begin{figure}[hbt]
\begin{center}
\includegraphics[width=8cm]{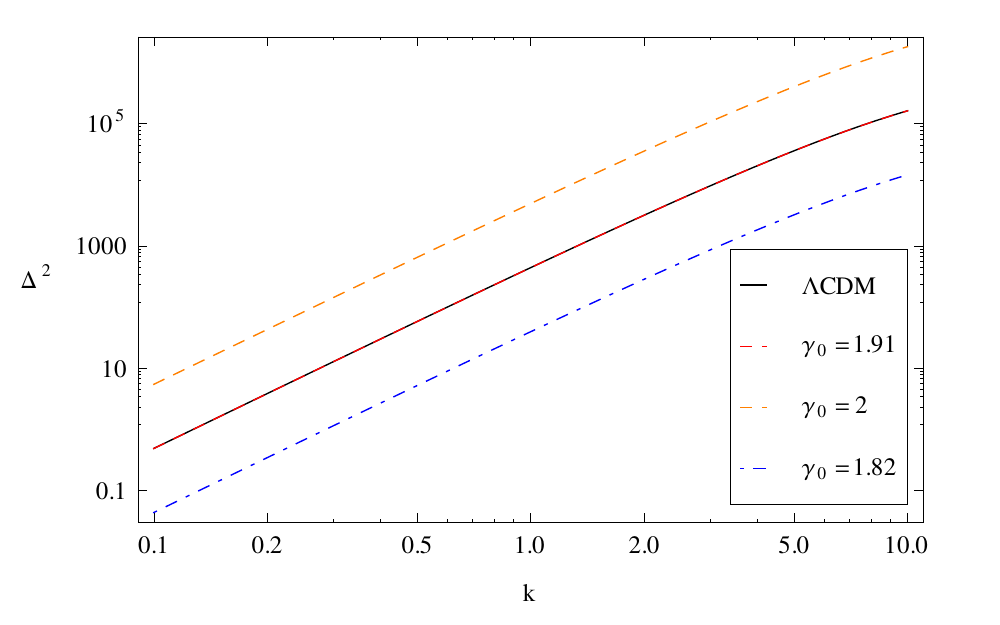}
\caption{\scriptsize{Plot of the dimensionless power spectrum as function of $k$ at $a=1$, for the complementary tachyon fields with $\ga_0=1.82$, $\ga_0=1.91$ and $\ga_0=2$.}}
\label{F3}
\end{center}
\end{figure}

\begin{figure}[hbt!]
\begin{center}
\includegraphics[width=7.5cm]{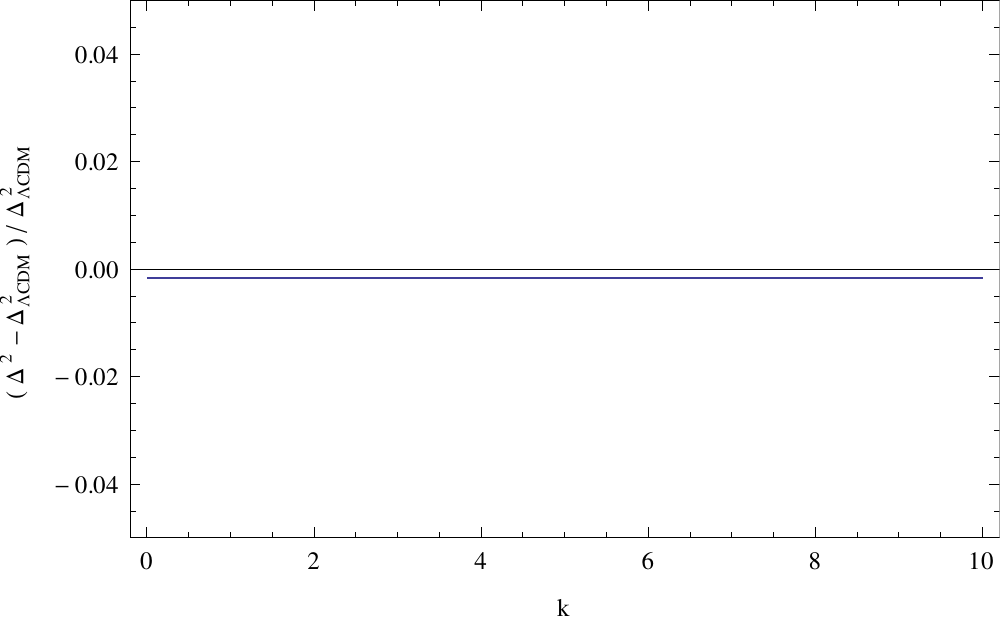}
\caption{\scriptsize{Plot of  the fractional difference of the power spectra of the complementary tachyon with $\ga_0=1.91$ from that of the $\Lambda$CDM model.}}
\label{F4}
\end{center}
\end{figure}

The Figure \ref{F3} describes the power spectrum for the complementary tachyon field for $\ga_0=1.82$, $\ga_0=1.91$ and $\ga_0=2$ and for the $\Lambda$CDM model. It is observed that the curve of the complementary tachyon field with $\ga_0=1.91$ coincides, has the same shape and amplitude, with the one of the concordance model. It can be seen from Figure \ref{F4}, that there is a fractional difference of the power spectra of the complementary tachyon with $\ga_0=1.91$ from that of the $\Lambda$CDM model is very small.

\vskip 0.2cm

From Figure \ref{F1} and  \cite{LuisTaq2}, it can be inferred that the models with $\ga_0=0.99$ and $\ga_0=1.01$ closely resemble the $\Lambda$CDM model. But when density perturbations are taken into account  Figure \ref{F2} shows that the two models with that barotropic index differ considerably. This result is physically relevant as it implies that the large structure formation that these models predict are far from observations. The only special case, as we indicated above, are the ones with barotropic index very close to $\gamma_0= 1.91$.

Our interpretation about the discrepancies given above is the following. For the fiducial model the speed of sound $c_s$ in (\ref{Eu}) is zero. The same happens with the tachyon model with $\gamma_0\to 1$, since the speed of sound is $c^2_s=1-\gamma_0$, so the equations look similar. However, this equation involves the factor (\ref{HLa}) which differs for both models. When $\gamma_0\to 1$ the tachyon factor is
$$
\frac{\dot{H}}{H^{2}}\to-\frac{3}{2}\frac{1}{1+a^{3\ga_0}}.
$$
Instead for the $\Lambda$CMD model this factor is  known to be
\begin{equation}\label{ivangrozny}
\frac{\dot{H}}{H^{2}}=-3\frac{1}{1+\alpha a^{3\ga_0}}
\end{equation}
where $\alpha$ has a value that depends of the density parameters of the model. By comparing both factors it is seen that there is a coefficient $1/2$ of discrepancy between the two models. This simple difference can introduce radical change of behavior, even in linear differential equations. To give an example consider an hypergeometric equation
$$
z(1-z)u''+[a-(b+c+1)z]u'-bc u=0.
$$
This equation is clearly linear. One of its elementary solutions is the hypergeometric function $F(a, b, c,z)$. This solution is convergent in the boundary $|z|=1$ when \cite{grady}
$$
\Re(b+c-a)\leq 0.
$$
Suppose that $b$ and $c$ have a fixed value, say $b+c=2$. If for instance $a=3$, then the condition above shows that $F(a,b,c,z)$ is convergent in  $|z|=1$. However, if we divide $a$ with a factor $2$, then $a=3/2$ and the condition above is not satisfied. Thus the solution $F(a,b,c,z)$ has a divergent behavior at the boundary  $|z|=1$. 

The example given above shows that a simple numerical factor in a linear equation can modify radically the behavior of the solutions. Unfortunately we do not know the exact solution of our equation (\ref{Eu}), but our numerical results are suggesting that a subtle effect of this type is the one which gives the discrepancy between the power spectrum of the concordance model and the tachyon model with $\gamma_0\to 1$. On the other hand, for $\gamma_0=2$ the factor (\ref{HLa}) becomes the same as for the concordance model, but the speed of sound $c_s$ is not zero and therefore the third term in Eq. (\ref{Eu}) do not vanish, thus the power spectrum is also different. Note that we did not even take into account the non trivial value of $\alpha$ in the factor (\ref{ivangrozny}), which also may generate discrepancies. The power spectrum is better approximated at some value between $\gamma_0\to 1$ and $\gamma_0=2$, which we found numerically to be $\gamma_0\sim 1.91$, as shown in Fig. \ref{F3}.

\section{Statefinder Diagnostic}
The statefinder diagnostic is a diagnos introduced in \cite{Sahni}, which allows to distinguish between several dark energy models. This diagnostic is constructed starting with the scale factor $a(t)$ and its derivatives up to the third order. The statefinder pair $\{s,r\}$ \cite{Tong} \cite{Pano} \cite{Ishi} is defined by
\begin{equation}
\label{Star}
r=\frac{\dddot{a}}{aH^{3}}=\frac{\ddot{H}}{H^3}-3q-2,  
\end{equation}
\begin{equation}
\label{Stas}
s=\frac{r-1}{3(q-1/2)},
\end{equation}
where $q$ is the deceleration parameter, defined by $q\equiv -\ddot{a}/aH^2$. The statefinder pair is a geometrical diagnostic in the sense that it is constructed directly from the space-time metric. 

In the situations in which the space time is a flat FRW universe, the standard $\Lambda$CDM model corresponds to a fixed point $\{r,s\}_{\Lambda CDM}=\{1,0\}$. This value is independent on the parameter of $\Lambda$CDM model and the redshift $z$. Hence, we apply the statefinder diagnostic to the tachionization model for the different barotropic index $\gamma_0$. By plotting the trajectories in the $r-s$ phase diagram, the distance of the model from the concordance model can be probed, as discussed in \cite{Alam}.

By use of Eqs. (\ref{HL}), (\ref{aL}) and its derivates, the deceleration parameter and Eq. (\ref{Star}) can be expressed as
\begin{equation}
\label{q}
q(z)=-\frac{3}{2}\left[ (\frac{2}{3\gamma_0}-1)+\frac{1}{(z+1)^{3\gamma_0}+1}\right], 
\end{equation}
\begin{equation}
\label{Star2}
r(z)=\frac{9}{2}\gamma_0^2\frac{(z+1)^{3\gamma_0}}{(z+1)^{3\gamma_0}+1}.
\end{equation}
The remaining parameter $s$ can be obtained by substituting equation (\ref{Star2}) in equation (\ref{Stas}). We avoid to rewrite it here.

\begin{figure}
\begin{center}
\includegraphics[width=5.5cm]{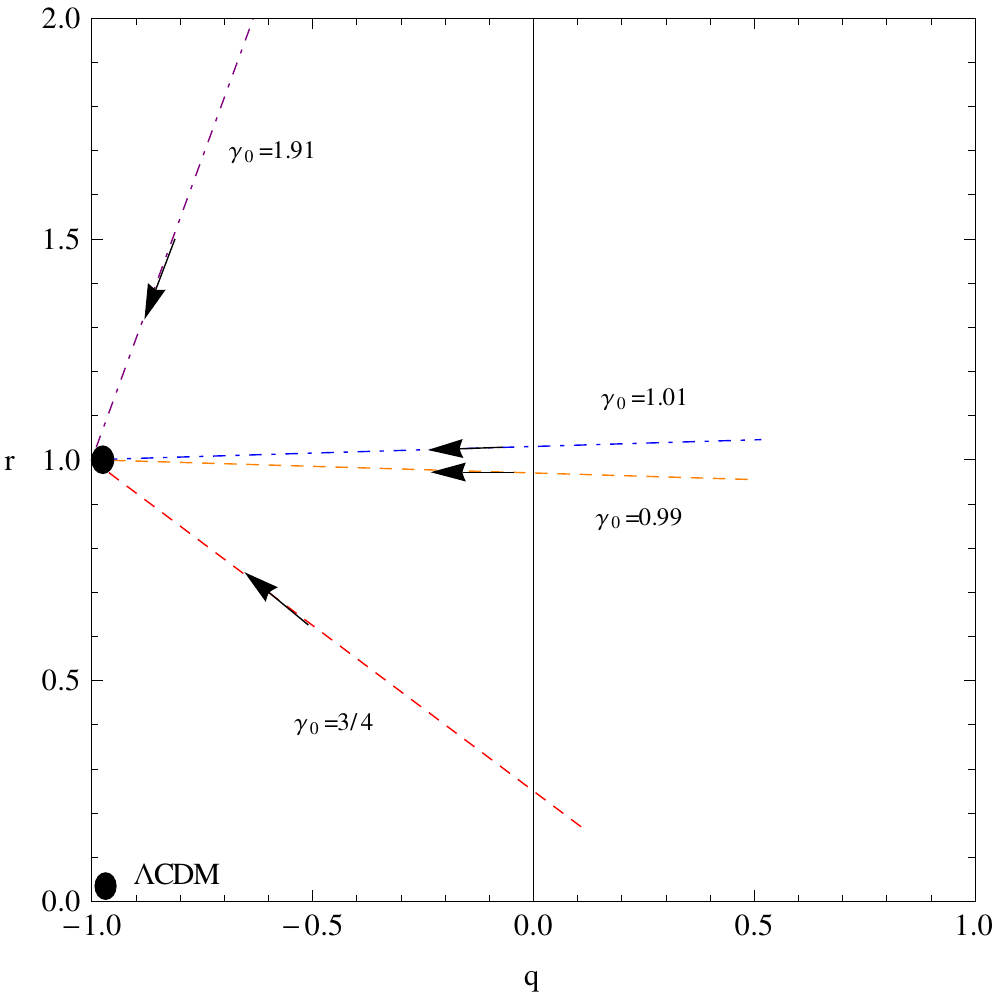}
\caption{\scriptsize{Trajectories in $\{r, q\}$ phase plan with $\gamma_0 = 3/4$, $\gamma_0 = 0.99$, $\gamma_0 = 1.01$ and $\gamma_0 = 1.91$. The black dot is the stable state of $\{r, q\}$ in the future.}}
\label{F5}
\end{center}
\end{figure}

\begin{figure}
\begin{center}
\includegraphics[width=5.5cm]{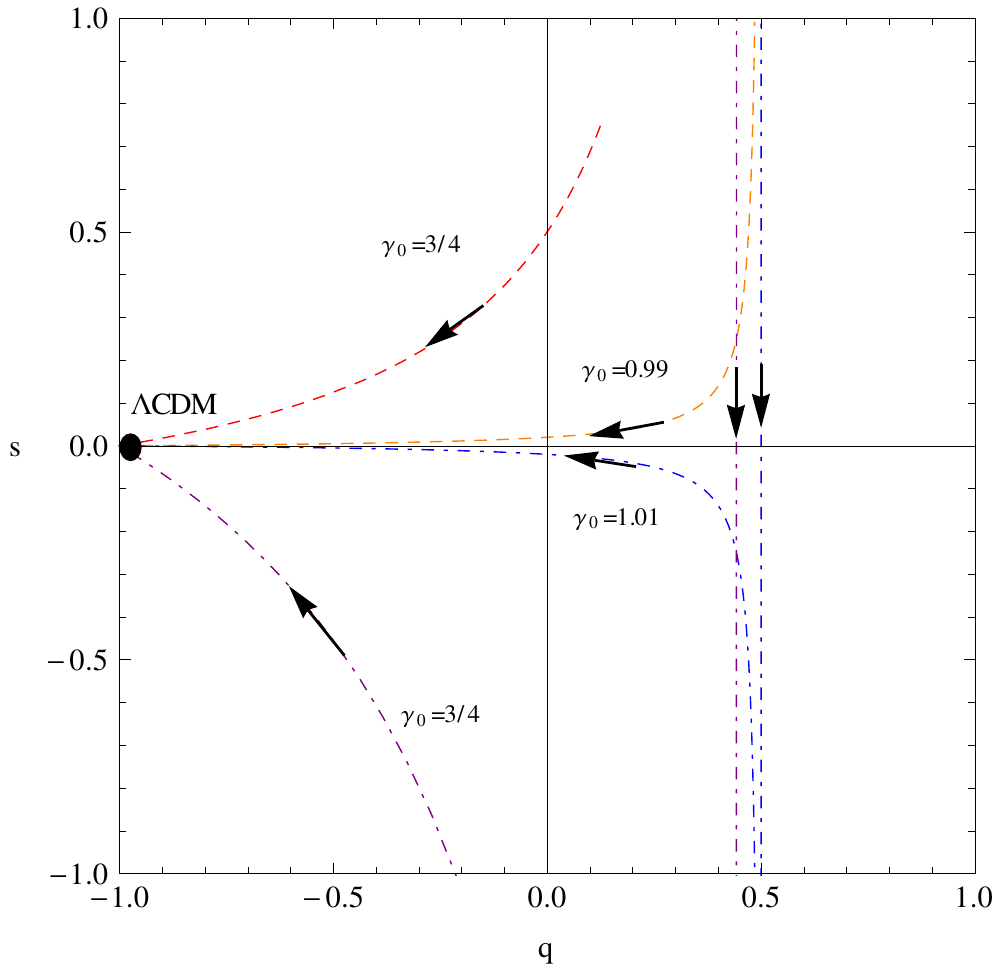}
\caption{\scriptsize{Trajectories in $\{s, q\}$ phase plan with $\gamma_0 = 3/4$, $\gamma_0 = 0.99$, $\gamma_0 = 1.01$ and $\gamma_0 = 1.91$. The black dot is the stable state of $\{s, q\}$ in the future.}}
\label{F6}
\end{center}
\end{figure}

\begin{figure}
\begin{center}
\includegraphics[width=5.5cm]{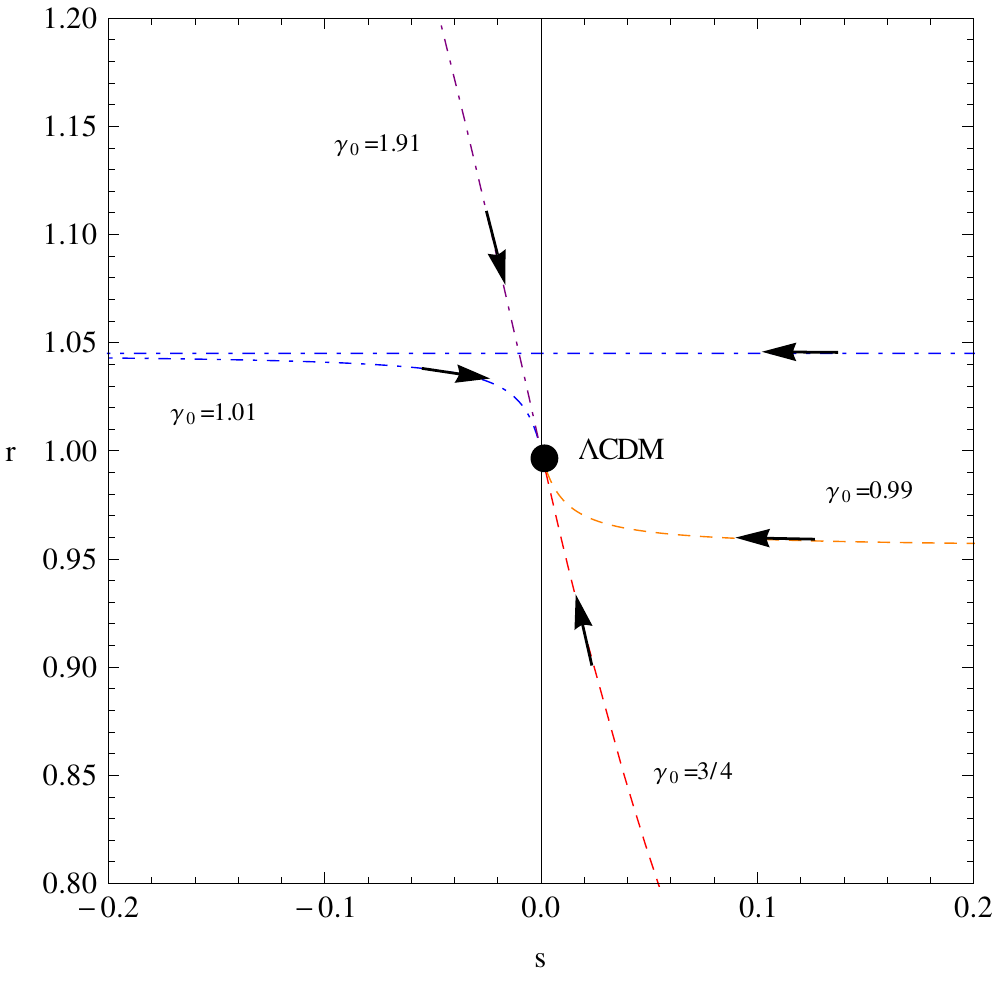}
\caption{\scriptsize{Trajectories in $\{r, s\}$ phase plan with $\gamma_0 = 3/4$, $\gamma_0 = 0.99$, $\gamma_0 = 1.01$ and $\gamma_0 = 1.91$. The black dot is the stable state of $\{r, s\}$ in the future.}}
\label{F7}
\end{center}
\end{figure}

Let us  analyze the parametric plots of the Statefinder diagnostic. The Fig. \ref{F5} shows the trajectories of $\{r, q\}$ phase plane for four different values of the parameter $\gamma_0$. This figure suggest that all the trajectories evolve to a fixed point in the future. The evolution for the trajectories with $\gamma_0=0.99$ and $\gamma_0=1.01$ (close to dust) is similar and overly far from the evolutions with $\gamma_0=3/4$ and $\gamma_0=1.91$. The phase plane for $\{s, q\}$ is showed in Figure \ref{F6}. Here also, the trajectories tend to a fixed point in the future, that coincides with the fixed point of the $\Lambda$CDM model (with $\{s,q\}=\{1,-1\}$). As expected, the evolution of the trajectories for the different barotropic index tend the one corresponding to a cosmological constant, so this model could explain the late time speed-up expansion of the universe. Finally, the $\{r, s\}$ diagram can be seen in Fig. \ref{F7}. In this phase plane also, the trajectories evolve to a stable point in future which is corresponding to concordance model.

In the Table \ref{StI}  the currents values of the parameters $q$, $r$ and $s$ are represented for the different values of the barotropic index analysed in the work, $\gamma_0=3/4$, $\gamma_0=0.99$, $\gamma_0=1.01$ and $\gamma_0=1.91$. The present-day value of the deceleration parameter $q(z=0)\in[-0.62;-0.56]$ as stated in the WMAP-9 report \cite{WMAP9}, is below form the ones found in this work for all the  barotropic index. In particular, for $\gamma_0=1.91$ the model is not in a acceleration phase, but goes to one in the future. The current values $r(z=0)$ and $s(z=0)$ in the model differ from $1$ and $0$ respectively, for all the cases analyzed, see Table \ref{StI}.   

\begin{center}
\begin{table}[ht!]
\centering
\scalebox{0.85}{\begin{tabular}{|l|l|l|l|}
  \hline
  \multicolumn{4}{|c|}{Tachyonization of the $\Lambda$CDM model} \\
  \hline
  $\gamma_0$ & $q(z=0)$ & $r(z=0)$ & $s(z=0)$ \\
  \hline
  $\gamma_0=3/4$ & -0.44 & 0.58 & 0.15  \\
  \hline
  $\gamma_0=0.99$ & -0.26 & 0.98 & 0.01  \\
  \hline
  $\gamma_0=1.01$ & -0.24 & 1.02 & -0.01  \\
  \hline
  $\gamma_0=1.91$ & 0.43 & 4.91 & -19.31 \\
  \hline
\end{tabular}}
\caption{\label{StI} The present values of the currents $q$, $r$ and $s$ parameters for different barotropic index.}
\end{table}
\end{center}

\section{Summaries and Conclusion}

In the present work the power spectrum for the tachyon models of \cite{LuisTaq2}, \cite{LuisTE} was investigated by numerical methods, by taking (\ref{CI}) as initial conditions.  The power spectrum of the standard and complementary tachyon was compared with the spectrum of the  fiducial $\Lambda$CDM model, since we know that the standard $\Lambda$CDM model is pretty much consistent to the current cosmological observations.
This was done by taking into account cosmological parameters which are in agreement with the ones obtained in WMAP9 \cite{WMAP9} and the Planck mission \cite{Planck2013}. It was found that both spectrums are quantitatively different and can be distinguished, even for small $\Lambda$ values. The power spectrum obtained strongly depends on the value of the barotropic index $\gamma_0$ of the tachyon in consideration, although the qualitative behaviour of the power spectrum curve is similar for all the cases. For instance, the power 
spectrum for the cases with $\gamma_0=3/4$ differ in several orders of magnitude from the one with $\gamma_0=0.99$. We have found that the tachyon model that better agrees with the $\Lambda$CMD is the one with barotropic index $\gamma_0=1,91$. This corresponds to a complementary tachyon. The curves for the power spectrum corresponding to $\gamma_0<1,91$ are strictly below the $\Lambda$CMD model, while the opposite holds for $1.91<\gamma_0<2$. 

In addition, the statefinder diagnostic for the minimal case has also been studied. The plot of the phase plan of the pairs $\{r, q\}$, $\{s, q\}$ and $\{r, s\}$ suggested that the trajectories of the all cases, for the first pair, tend to the state which has a non vanishing distance from $\Lambda$CDM model. In contrast, the last two pairs reaches a stable state which is corresponding to the concordance scenario, so the tachyonization of the $\Lambda$CDM model is capabie to explain the late time accelerating phase of the universe expansion. In the Table \ref{StI}, it is shown the deceleration parameter $q$ for different values of the barotropic index, and it is found that for $\gamma_0=1.91$ the $q(0)=0.43$, which implies that the universe is decelerating instead of accelerating. An odd observation is that this value of $\gamma_0$ gives a similar power spectrum that the fiducial model, although a decelerating universe is ruled out by the observations. On the other hand, the values of $\gamma_0\sim 1$ which can result to similar expansion history as $\Lambda$CDM model, as the Figure \ref{F1} shows, have several orders of magnitude difference in amplitude of the power spectrum in comparison to the fiducial model. So it seems that the assumed Tachyonic model cannot explain all observations simultaneously.

Our work did not considered inflationary aspects of this tachyon fields, and it may be of interest to study them. We leave this for a future investigation.

\acknowledgments
We ackowledge the anonymous referee for some comments about the first version of this paper which improved the description of the results. The authors are supported by the CONICET, Argentina.





\end{document}